\documentclass[11pt]{article}

\usepackage[fleqn]{amsmath}
\usepackage{amsfonts}
\usepackage{amssymb}
\usepackage{epsfig}
\usepackage{geometry}
\usepackage{ctable}

\DeclareMathOperator{\sign}{sgn}

\newcommand{\pdt}{{\partial_t^{\phantom{0}}}}
\newcommand{\pdtsq}{{\partial_t^2}}

\newcommand{\NullV}{\boldsymbol{0}}

\newcommand{\ba}{\mathbf{a}}
\newcommand{\bB}{\mathbf{B}}

\newcommand{\bD}{\mathbf{D}}
\newcommand{\bE}{\mathbf{E}}

\newcommand{\bH}{\mathbf{H}}

\newcommand{\pB}{\boldsymbol{p}}
\newcommand{\qB}{\boldsymbol{q}}
\newcommand{\vB}{\boldsymbol{v}}

\newcommand{\drm}{\mathrm{d}}
\newcommand{\Drm}{\mathrm{D}}

\newcommand{\Ddt}{\frac{\drm\phantom{s}}{\drm t}}

\newcommand{\pddt}{\frac{\partial\phantom{t}}{\partial t}}
\newcommand{\tpddt}{{\textstyle{\frac{\partial\phantom{t}}{\partial t}}}}

\newcommand{\refeq}[1]{(\ref{#1})}

\newcommand{\mbare}{m_{\text{b}}}

\newcommand{\vect}[1] {\boldsymbol{{ #1}} }

\newcommand{\qv}[1]{{\textbf{\textrm{#1}}}}

\newcommand{\tenseur}[1]{{\textbf{\textsf{#1}}}}

\newcommand{\Rset}{\mathbb{R}}
\newcommand{\Sset}{\mathbb{S}}

\newcommand{\ID}{{\boldsymbol{I}_{3\times3}^{}}} 

\newcommand{\FQ}{\tenseur{F}}           
\newcommand{\gQ}{\tenseur{g}}           
\newcommand{\etaQ}{\boldsymbol{\eta}}   
\newcommand{\RQ}{\tenseur{R}}           
\newcommand{\TQ}{\tenseur{T}}           

\newcommand{\fQ}{\qv{f}}        	
\newcommand{\qQ}{\qv{q}}                
\newcommand{\uQ}{\qv{u}}                

\newcommand{\aV}{\vect{a}}              

\newcommand{\fV}{\vect{f}}              
\newcommand{\nV}{{\vect{n}}}		
\newcommand{\pV}{{\vect{p}}}            
\newcommand{\qV}{{\vect{q}}}            
\newcommand{\sV}{{\vect{s}}}            
\newcommand{\vV}{{\vect{v}}}            

\newcommand{\PiV}{\boldsymbol{\Pi}}
\newcommand{\nab}{\vect{\nabla}}

\newcommand{\abs}[1]{\big\vert #1 \big\vert}

\newcommand{\ul}[1]{\underline {#1} }

\renewcommand{\leq}{\leqslant}
\renewcommand{\geq}{\geqslant}

\newcommand{\crprd}{{\boldsymbol\times}}

\reversemarginpar

\numberwithin{equation}{section}

\newtheorem{thm}{Theorem}[section]

\newtheorem{prp}[thm]{Proposition}

\newtheorem{rem}[thm]{Remark}

\begin{document}

%
\title{{ The Einstein-Infeld-Hoffmann Legacy\\ in Mathematical Relativity\\ 
          \large{\sc{I: The Classical Motion of Charged Point Particles}}}\footnote{Rapport by the first author given 
                        at the 15th Marcel Grossmann Meeting on General Relativity, 
                        Univ. Roma 1 (La Sapienza), Rome, Italy, July 3, 2018. Final revision: August 15, 2019.}}
\author{\textbf{M. K.-H. Kiessling and A. S. Tahvildar-Zadeh}\\
                \small{Department of Mathematics, Rutgers University}\\
                \small{110 Frelinghuysen Rd., Piscataway, NJ 08854, USA}\\
                {Version of August 15, 2019}\\ 
\textrm{\small Typeset with \LaTeX\ on: }}
\vspace{-1truecm}
\maketitle

\thispagestyle{empty}

\vspace{-1truecm}
\begin{abstract}
\noindent 
 Einstein, Infeld, and Hoffmann (EIH) claimed that the field equations of general relativity theory alone imply
the equations of motion of neutral matter particles, viewed as point singularities in space-like slices of spacetime;
they also claimed that they had generalized their results to charged point singularities.
 While their analysis falls apart upon closer scrutiny, the key idea merits our attention.
 This rapport identifies necessary conditions for a well-defined general-relativistic joint initial value problem 
of $N$ classical point charges and their electromagnetic and gravitational fields.
 Among them, in particular, is the requirement that the electromagnetic vacuum law guarantees a 
finite field energy-momentum of a point charge.
 This disqualifies the Maxwell(--Lorentz) law used by EIH.
 On the positive side, if the electromagnetic vacuum law of Bopp, Land\'e--Thomas, and Podolsky (BLTP) is used, and
the singularities equipped with a non-zero bare rest mass, then a joint initial value problem can be formulated in 
the spirit of the EIH proposal, and shown to be locally well-posed --- \emph{in the special-relativistic zero-$G$ limit}.
 With gravitational coupling (i.e. $G>0$), though, changing Maxwell's into the BLTP law and assigning a bare rest mass
to the singularities is by itself not sufficient to obtain even a merely well-defined joint initial value problem:
the gravitational coupling also needs to be changed, conceivably in the manner of Jordan and Brans--Dicke.
\end{abstract}

\vfill
\hrule
\smallskip

\copyright(2019) \small{The authors. Reproduction of this preprint, in its entirety, is permitted for

non-commercial purposes only.}

\newpage



                \section{Brief History and State of Affairs}

\begin{quote}{\footnotesize ``I am plaguing myself with the derivation of the equations of motion of material points,
 conceived of as singularities [in the gravitational field], from the equations of general relativity.'' 
\hfill\hfill \textbf{Albert Einstein}, in a letter to \textbf{Max Born} on Dec. 4, 1926.}
\end{quote}

 We don't know when \textbf{Einstein} first conceived of the notion of point particles as singularities in relativistic 
fields,\footnote{In 1909 he remarked that ``light quanta'' might be point singularities in
  ``a field,'' their motion being guided by the electromagnetic field. 
  See part II of our rapport.\vspace{-.5truecm}}
but his letter to Max Born makes it plain that by the end of 1926 his ideas had matured to the point where he pursued a 
dynamical theory for such point singularities, expecting that their law of motion could be extracted from his gravitational 
field equations. 
  Already a month later \textbf{Einstein} \&\ \textbf{Grommer} announced that 
``the law of motion is completely determined by the field equations, though shown in this work only for the case of equilibrium.''
  In that paper \cite{EinsteinGrommer} the case of a static, spherically symmetric spacetime with a single time-like singularity was studied. 
 The truly dynamical many-body problem was treated a decade later by \textbf{Einstein, Infeld}, and \textbf{Hoffmann}
in their famous paper \cite{EinsteinInfeldHoffmann}, with follow-ups in \cite{EinsteinInfeldA} and \cite{EinsteinInfeldB}.
 They argued explicitly that the field equations of general relativity theory alone determine the equations of motion 
of neutral matter particles, viewed as point singularities in space-like slices of spacetime. 
 They also claimed that they had generalized their results to charged point-singularities, with the details written up
in a set of notes deposited with the secretary of the IAS. 
 In 1941 the motion of charged point-singularities was revisited by Infeld's student \textbf{P. R. Wallace},
who presented the details of the calculations in \cite{Wallace}.

 Here is the gist of the \textbf{Einstein-Infeld-Hoffmann} argument (modern terminology):
\begin{itemize}
\item 
Suppose you have a four-dimensional, time-oriented, asymptotically flat {electromagnetic \textbf{Lorentz} spacetime} $\mathcal{M}^{1,3}$ with 
$N$ charged, {time-like singularities} of infinite extent, presentable as {a graph over $\Rset^{1,3}\setminus \{N$ time-like world-lines$\}$.}
\item Away from the singularities the spacetime structure obeys \textbf{Einstein}'s equations
\begin{equation}\label{EINSTEINeq}
 \RQ-{\textstyle\frac12}R\gQ 
= {\textstyle\frac{8\pi G}{c^4}}\TQ[\FQ,\gQ],
\end{equation}
where $\TQ[\FQ,\gQ]$ is the energy-momentum-stress tensor of the electromagnetic vacuum field $\FQ$,
satisfying \textbf{Maxwell}'s field equations in vacuum,
\begin{equation}
\boldsymbol{d}\FQ = \boldsymbol{0}\qquad \&\qquad 
\boldsymbol{d*}\!\FQ = \boldsymbol{0}.\vspace{-5pt}
\end{equation}
\item
 The twice contracted second \textbf{Bianchi} identity implies energy-momentum conservation:
\begin{equation}
\boldsymbol{\nabla\cdot}\left(\RQ-{\textstyle\frac12}R\gQ \right) = \boldsymbol{0} 
 \implies {\boldsymbol{\nabla\cdot}\TQ[\FQ,\gQ] = \boldsymbol{0} }.\vspace{-5pt}
\end{equation}
\item ``Massive, charged'' singularities are associated with field ``fluxes,'' and thus
\begin{equation}\label{divTnullPLUSflux}
 \boxed{\boldsymbol{\nabla\cdot}\TQ[\FQ,\gQ] = \boldsymbol{0} \quad\&\quad\mbox{flux\ conditions}}
\implies
{\mbox{law\ of\ the\ time-like\ singularities}}.\vspace{-5pt}
\end{equation}
\end{itemize}
 The main bullet point is of course the last one.
 Here are \textbf{EIH} in their own words (p.66): 
``{It is shown that \ul{for two-dimensional} [closed] \ul{spatial surfaces containing singularities}, 
certain \ul{surface} \ul{integral conditions} are valid which \ul{determine the motion}.}'' [Emphasis ours.]

 Unfortunately, despite its publication in the Annals of Mathematics, the 1938 \textbf{EIH} paper is not only not rigorous, 
it contains questionable technical assumptions and serious blunders.
 Some were addressed in \cite{EinsteinInfeldB}, yet \emph{their main conclusions turn out to be false}. 
 All the same, the core idea merits further inspection.
 
 The issue is how to correctly handle singularities.
 \textbf{EIH} state (p.66): ``{By means of a {new method of approximation}, specially suited to the treatment of 
quasi-stationary fields, the {gravitational field due to moving particles is determined}.}''
 This approximation method assumes that the particles are moving \emph{slowly} and the 
field strengths (as seen in a \textbf{Lorentz} frame of \textbf{Minkowski} space, in which the particles move slowly) are \emph{weak}. 
 As a consequence, one has to choose the radii of the closed surfaces sufficiently large to satisfy the 
weak-field assumption needed for the convergence of the expansion (which was not shown). 
 Yet, on p. 92 one reads: ``It is most convenient to take definite, {infinitesimally small spheres 
whose centers are at the singularities}, ...'' which clearly violates their weak-field condition. 
 Indeed, \textbf{EIH} realize that: ``... in this case {infinities of the types}
\begin{equation}
\lim const. / r^n,\qquad\mbox{$n$\ a positive integer},\qquad r \to 0
\end{equation}
{can occur} in the values of the partial integrals,''
but then commit a major blunder by stating (\textbf{EIH}, p. 92): ``Since {these must cancel}, however, in the final result, 
{we may merely ignore them} throughout the calculation of the surface integrals.'' 

\centerline{\emph{Alas, the infinities do {not} cancel!}}

 In the same year \textbf{P. A. M. Dirac} invented \emph{negative infinite bare mass renormalization} to 
handle those infinities, in the simpler special-relativistic purely electrodynamical setting \cite{DiracA}.
 For an electron with positive ``observable mass'' $m_{\mbox{\tiny{obs}}}$ and charge $-e$ he obtained the equation of motion
(in \textbf{Misner--Thorne--Wheeler} notation)
\begin{equation}\label{AbrLorDiracEQ}
\textstyle
m_{\mbox{\tiny{obs}}}
\frac{\drm^2}{\drm\tau^2}\qQ
= 
\fQ^{\mbox{\tiny{ext}}} + \fQ^{\mbox{\tiny{\sc{Laue}}}} ,
\end{equation}
where 
\begin{equation}\label{Lf}
\textstyle \fQ^{\mbox{\tiny{ext}}} = -\frac{e}{c}\FQ^{\mbox{\tiny{ext}}}(\qQ)\cdot\frac{\drm}{\drm\tau}\qQ
\end{equation}
is a \textbf{Lorentz Minkowski}-force due to ``external sources,'' 
\begin{equation}\label{vLMf}
\textstyle
\fQ^{\mbox{\tiny{\sc{Laue}}}} = 
\frac{2e^2}{3c^3} 
\left(\gQ +\frac{1}{c^2}\frac{\drm}{\drm\tau}\qQ \otimes\frac{\drm}{\drm\tau}\qQ\right)\cdot
{\textstyle\frac{\drm^3}{\drm\tau^3}\qQ }  
\end{equation}
is \textbf{von Laue}'s radiation-reaction \textbf{Minkowski}-force of the electron, and 
\begin{equation}
{\textstyle m_{\mbox{\tiny{obs}}} = \lim_{r\downarrow 0}\left({\mbare} (r) +\frac{e^2}{2c^2}\frac1r\right)}
\end{equation}
defines ${\mbare}(r)$. [N.B.: ${\mbare}(r)\downarrow -\infty$ as $r\downarrow 0$].
 Here, $r$ is the {radius of a sphere} in the instantaneous rest-frame of the electron, centered at the electron,
which plays the role of the surfaces containing singularities invoked by \textbf{EIH}.
 
 As is well-known, such mass-renormalization computations have become the template for the much more elaborate ---
and quite successful --- renormalization group computations in quantum electrodynamics (and more generally, quantum field theory).
 Be that as it may, \textbf{Dirac} himself later in life became very dissatisfied with this approach, and so are we.

 First of all, supposing a point electron has a bare mass, then how could it possibly depend on the radius $r$ of a sphere over
which a theoretical physicist averages the fields?
   
 Second, the third proper-time derivative 
featuring in the \textbf{von Laue Minkowski}-force means that \refeq{AbrLorDiracEQ} is a 
third-order ODE for the position of the particle as a function of proper time, requiring vector initial data for position, 
velocity, and acceleration. 
 Yet a classical theory of point particle motion should only involve initial data for position and velocity.
 
 In 1951 \textbf{Lev Landau} \&\ \textbf{Eugenii Lifshitz}  addressed the $\dddot\qQ$ problem as follows:
\begin{itemize}
\item
{Test particle theory works well for many practical purposes}.
\item
In such situations \textbf{von Laue}'s  $\dddot\qQ$ {force term must be a small perturbation} of $\fQ^{\mbox{\tiny{ext}}}$.
\item
{Compute} $\dddot\qQ$ perturbatively: take the 
proper-time derivative of the test-particle law, and obtain
\begin{equation}\label{dddq}
\textstyle
\frac{\drm^3}{\drm\tau^3}\qQ            
\approx -\frac{e}{m_{\mbox{\tiny{obs}}}c}\frac{\drm}{\drm\tau}
\left(\FQ^{\mbox{\tiny{ext}}}(\qQ)\cdot\frac{\drm}{\drm\tau}\qQ\right).
\end{equation}\vspace{-10pt}
\item 
{The right-hand side depends only on} $\qQ$, $\dot\qQ$, $\ddot\qQ$. 
Substitute it for $\frac{\drm^3}{\drm\tau^3}\qQ$ in \refeq{vLMf}.
\end{itemize}
 R.h.s.\refeq{vLMf} with r.h.s.\refeq{dddq} substituted for
$\frac{\drm^3}{\drm\tau^3}\qQ$ will be called the \textbf{Landau--Lifshitz Minkowski}-force of radiation-reaction, denoted 
$\fQ^{\mbox{\tiny{LL}}}$. 
 Equation \refeq{AbrLorDiracEQ} with $\fQ^{\mbox{\tiny{\sc{Laue}}}}$ replaced by $\fQ^{\mbox{\tiny{\sc{LL}}}}$ is known
as the \textbf{Landau--Lifshitz} equation of motion for the electron.
 It seems to work quite well for practical purposes in which $\FQ^{\mbox{\tiny{ext}}}$ can be approximated by some
smooth field tensor, on time scales beyond the one where test particle theory works well, but not arbitrarily far beyond
\cite{Spohn}.
 However \cite{DeckertHartenstein}, as soon as $\FQ^{\mbox{\tiny{ext}}}$ is taken to be the field generated by all other particles 
the \textbf{Landau--Lifschitz} equation of motion is typically well-posed only until the moment that 
a point charge meets the forward initial light cone of another point charge, a ludicrously short time span!

 \textbf{Dirac}'s idea of infinite negative bare mass renormalization and \textbf{Landau--Lifshitz}'s perturbative treatment
of the \textbf{von Laue} radiation-reaction \textbf{Minkowski}-force have become standard ingredients also in general-relativistic
treatments of charged point particle motion. 
 Thus, for a point electron moving in an externally given curved background, \textbf{Eric Poisson, Adam Pound}, \&\ \textbf{Ian Vega} 
in their review  \cite{PPV} present the following equations of motion:
\begin{equation}\label{PPVeqOFmot}
\textstyle
m_{\mbox{\tiny{obs}}} \frac{\Drm}{\drm\tau}\uQ
= 
\fQ^{\mbox{\tiny{ext}}}  + \fQ^{\mbox{\tiny{\sc{Laue}}}} + \fQ^{\mbox{\tiny{tail}}} ,
\end{equation}
where $\uQ := \frac{\drm}{\drm\tau}\qQ$ and 
$\frac{\Drm}{\drm\tau}\uQ :=
 \frac{\drm}{\drm\tau}\uQ + \boldsymbol{\Gamma}^{\mbox{\tiny{ext}}}\left(\uQ,\uQ\right)$,
and $\fQ^{\mbox{\tiny{ext}}} = -\frac{e}{c}\FQ^{\mbox{\tiny{ext}}}(\qQ)\cdot\uQ$ as before, but now
\begin{equation}\label{fLAUEagain}
\textstyle
\fQ^{\mbox{\tiny{\sc{Laue}}}} = 
\frac{2}{3} e^2
\left(\gQ +\frac{1}{c^2}\uQ \otimes\uQ\right)\cdot
\left(\frac16 \RQ^{\mbox{\tiny{ext}}} \cdot \frac{1}{c}\uQ  
+ \frac{1}{c^3} {\frac{\Drm^2}{\drm\tau^2}\uQ }\right)\vspace{-5pt}
\end{equation}
with
\begin{equation}\label{dddqAGAIN}
\textstyle
{\frac{\Drm^2}{\drm\tau^2}\uQ }
\approx - \frac{e}{m_{\mbox{\tiny{obs}}}c}\frac{\Drm}{\drm\tau}
\left(\FQ^{\mbox{\tiny{ext}}}(\qQ)\cdot\uQ\right),
\end{equation}
and
\begin{equation}\label{fTAIL}
\fQ^{\mbox{\tiny{tail}}} = 2 e^2
\int_{-\infty}^\tau 
\textstyle
\boldsymbol{H}^{\mbox{\tiny{ret}}}(\qQ(\tau),\qQ(\tau'))\cdot
\uQ(\tau')\drm\tau'\cdot\uQ(\tau),
\end{equation}
where $\boldsymbol{H}^{\mbox{\tiny{ret}}}(\qQ(\tau),\qQ(\tau'))$ is a retarded type of Green function for the electromagnetic
field tensor in curved spacetime. 
 Equation \refeq{PPVeqOFmot} does not yet include gravitational radiation-reaction, which
\textbf{Poisson, Pound}, \&\ \textbf{Vega} discuss also, in particular the approaches
of \textbf{Quinn} \&\ \textbf{Wald} and \textbf{Detweiler} \&\ \textbf{Whiting} (see \cite{PPV}), but we don't need to go 
there because \refeq{PPVeqOFmot} already displays a major problem due to the so-called tail force term:

\noindent
 \emph{Equation \refeq{PPVeqOFmot}, even with r.h.s.\refeq{dddqAGAIN} substituted for} ${\frac{\Drm^2}{\drm\tau^2}\uQ}$ \emph{at
r.h.s.\refeq{fLAUEagain}, does not pose a second-order initial value problem for the position of the point electron
but instead requires the input of the entire past history of the motion!}
 
 One can try to extricate oneself from this dilemma by once again having recourse to a \textbf{Landau--Lifshitz}-type 
perturbation argument: the tail force, also a radiation-reaction term, must be small in situations where test particle theory works
well. 
 In this case,  backward from the initial instant (say at $\tau=0$) 
one can approximately replace $\qQ(\tau')$ and $\uQ(\tau')$ in the integrand by the pertinent expressions computed from test particle 
theory, with particle data for position and velocity given, and then treat this
so-modified equation as a second-order integro-differential equation from the initial instant on forward, with 
$\qQ(\tau')$ and $\uQ(\tau')$ in the integrand for $\tau'>0$ no longer test-particle expressions.
 This set of ``effective equations of motion'' may work well in many practical situations.

 Yet from a mathematical relativity point of view this state of affairs is very unsatisfactory, both technically (being 
non-rigorous) and conceptually (involving heuristic but arbitrary arguments).

                \section{Rigorous Approach}

 In the following we report on recent rigorous advances in formulating a joint initial value problem for classical charged point particles
and the electromagnetic and gravitational fields they generate, with the
key idea of the 1938 \textbf{EIH} paper, as outlined on p.1 of this rapport, serving as our point of departure.
 To avoid the mistakes made by \textbf{EIH}, we
inquire into necessary conditions on the energy-momentum-stress tensor which allow one to extract a 
law of motion associated with the time-like singularities from an equation like \refeq{divTnullPLUSflux} without 
invoking infinite mass renormalization, nor arbitrary averaging over some neighborhood of an a-priori ill-defined force field,
as done in \cite{PPV}.
 For the simpler special-relativistic zero-gravity limit (cf. \cite{miki}, \cite{mikishadi})
 we even state a well-posedness theorem, so this case is treated first.

                \subsection{The zero-$G$ Limit}
%
                \subsubsection{Time-like particle world-lines in Minkowski spacetime}

 In the limit $G\downarrow 0$, \refeq{EINSTEINeq} is solved by 
$\mathcal{M}^{1,3} = \Rset^{1,3}\setminus \{N$ time-like world-lines$\}$,
with metric $\gQ=\etaQ$ away from the world-lines. 
 The question is which conditions on $\TQ$ lead, in a mathematically clean way,
to the \textbf{EIH}-type conclusion
\begin{equation}\label{divTnullPLUSfluxAGAIN}
 \boxed{\boldsymbol{\nabla\cdot}\TQ[\FQ,\etaQ] = \boldsymbol{0} \quad\&\quad\mbox{flux\ conditions}}
\implies
{\mbox{law\ of\ the\ time-like\ world-lines}}.\vspace{-5pt}
\end{equation}
 To answer this question, we extend $\mathcal{M}^{1,3}$ continuously to $\Rset^{1,3}$ (by adding the time-like world-lines), 
and switch to a distributional formulation. 
 We can formulate everything for $N$ charged, massive, {time-like world-lines}, but for simplicity  we set $N=1$ in the following.

 Since the issue is the formulation of a well-posed initial value problem, we also choose an arbitrary Lorentz frame, 
with space vector $\sV\in \Rset^3$ and time $t\in\Rset$. 
 Then the space part of l.h.s.\refeq{divTnullPLUSfluxAGAIN} becomes the
\emph{local conservation law for the total momentum}, 
\begin{equation}
\textstyle
\pddt \PiV(t,\sV) + \nabla\cdot T (t,\sV)
= \label{eq:MOMconservation}
\NullV,
\end{equation}
where it is \emph{postulated} (compatible with ``minimal coupling'') that the total momentum vector-density
\begin{equation}\label{eq:totalMOMENTUMdef}
\PiV(t,\sV):=\PiV^{\mbox{\tiny{field}}}(t,\sV) + \PiV^{\mbox{\tiny{charge}}}(t,\sV),
\end{equation}
with $\PiV^{\mbox{\tiny{field}}}(t,\sV)$ 
the contribution from the field and $\PiV^{\mbox{\tiny{charge}}}(t,\sV)$ (a distribution) the usual
contribution from the point charge, which must be assigned a non-vanishing bare rest mass $\mbare$.
 Similarly, it is \emph{postulated} that the symmetric total stress tensor is 
\begin{equation}\label{eq:totalSTRESSdef}
T(t,\sV):=T^{\mbox{\tiny{field}}}(t,\sV)+T^{\mbox{\tiny{charge}}}(t,\sV),
\end{equation}
with $T^{\mbox{\tiny{charge}}}(t,\sV)$ the usual stress tensor of the point 
particle, and $T^{\mbox{\tiny{field}}}(t,\sV)$ that of the field (except for our unconventional choice of sign here).

 Incidentally, energy conservation follows as a corollary from momentum conservation.

 Next, we integrate \refeq{eq:MOMconservation} over all $\sV\in\Rset^3$, at $t\in\Rset$.
 This yields the balance law of momentum exchange
\begin{equation}\label{eq:momISconserved}
\textstyle
\frac{\drm}{\drm t} \pV(t) 
=  - \frac{\drm}{\drm t} \displaystyle\int_{\Rset^3}\PiV^{\mbox{\tiny{field}}}(t,\sV) d^3{s}.
\end{equation}
 Clearly, for \refeq{eq:momISconserved} to make sense, the field momentum vector-density $\PiV^{\mbox{\tiny{field}}}(t,\sV)$
has to be integrable over $\Rset^3$, and this integral differentiable in time.
 This rules out the \textbf{Maxwell--Lorentz} field equations, but leaves other options, notably the 
\textbf{\uppercase{M}{\tiny\sc{axwell}}--\uppercase{B}{\tiny\sc{orn}}--\uppercase{I}{\tiny\sc{nfeld}}} (\textbf{MBI}) and
\textbf{\uppercase{M}{\tiny\sc{axwell}}--\uppercase{B}{\tiny\sc{opp}}--\uppercase{L}{\tiny\sc{and\'e}}--\uppercase{T}{\tiny\sc{homas}}--\uppercase{P}{\tiny\sc{odolsky}}}
(\textbf{MBLTP}) field equations (see below).

 Comparing equation \refeq{eq:momISconserved} with \textbf{Newton}'s law for the rate of change of momentum,
\begin{equation}
\textstyle
\frac{\drm}{\drm t} \pV(t) = \fV(t),
\end{equation}
where $\fV(t)$ is the force acting on the particle at time $t$, it is clear that the force $\fV(t)$ needs to be 
extracted from r.h.s.\refeq{eq:momISconserved}. 
 Since the particle's bare momentum $\pV(t)$ is given in terms of its bare mass $\mbare$ and velocity $\vV(t)=\Ddt\qV(t)$
by the \textbf{Einstein-Lorentz-Poincar\'e} law
\begin{equation}
\pV(t) : = \label{eq:MOMinTERMSofVELO}
\mbare\frac{\vV(t)}{\sqrt{1 -\frac{1}{c^2}|\vV(t)|^2}},
\end{equation}
the expression for the force also has to be compatible with the requirements of a second-order initial value problem for the
position of the point particle!
 Thus, beside the existence of the 
time derivative of the space integral over the field momentum vector-density, it is important
that the result involves, initially, only the initial electromagnetic fields and the initial data for position and velocity
of the point particle and, at later times $t>0$, at most the history of position, velocity, and acceleration of the particle, 
and of the fields, from the initial instant on, yet not beyond $t$.
 Whenever this is possible we obtain a well-defined joint initial value problem for field and particle, which may or may not
be well-posed.
  
 We were able to explicitly extract a well-defined force on the point charge from the \textbf{MBLTP}
field equations, and to prove that the resulting joint initial value problem for charge and field is well-posed. 
 We expect this to be feasible also for the \textbf{MBI} field equations, but their formidable nonlinearity makes
rigorous progress a slow process.

 Common to all these classical systems of electromagnetic field equations are the 
\emph{pre-metric {\bf{Maxwell--Lorentz}} field equations},  viz. the {evolution equations} 
\begin{alignat}{1}
\pdt{\bB(t,\sV)}
&= 
-c\nab\times\bE(t,\sV) \, \vspace{-5pt}
\\ 
\pdt{\bD(t,\sV)}
&=
+c\nab\times\bH(t,\sV)  + 4\pi e \dot\qV(t)\delta_{\qV(t)}(s)
\end{alignat}
and the constraint equations 
\begin{alignat}{1}
\textstyle
\nab\cdot \bB(t,\sV)  
&= 
0\, \vspace{-5pt}
\\ 
\nab\cdot\bD(t,\sV)  
&=
- 4\pi e \delta_{\qV(t)}(s)\,
\end{alignat}
for the $\bB$, $\bD$ {fields}. 
 They differ in the \emph{Electromagnetic Vacuum Law}: $(\bB,\bD)\leftrightarrow(\bH,\bE)$.
 The \textbf{Born-Infeld} law \cite{BornInfeldBb} reads
\begin{alignat}{1}
&\bH=\frac{\bB - \frac{1}{b^2}\bD\times(\bD\times\bB)}{\sqrt{1+\frac{1}{b^2}(|\bB|^2+|\bD|^2) +\frac{1}{b^4}|\bB\times\bD|^2}} \\
&\bE=\frac{\bD - \frac{1}{b^2}\bB\times(\bB\times\bD)}{\sqrt{1+\frac{1}{b^2}(|\bB|^2+|\bD|^2) +\frac{1}{b^4}|\bB\times\bD|^2}}
\end{alignat}
The \textbf{Bopp--Land\'e--Thomas--Podolsky} law 
\cite{Bopp}, \cite{Lande}, \cite{LandeThomas}, \cite{Podolsky} reads 
\begin{alignat}{1}
       \bH(t,\sV)  
&= 
       \left(1  + \varkappa^{-2}\square\,\right) \bB(t,\sV) 
\\
        \bD(t,\sV) 
&=
        \left(1  + \varkappa^{-2}\square\,\right) \bE(t,\sV) \, .
\end{alignat}
(N.B.: $\square := c^{-2}\pdtsq -\Delta$.)
When $b\to\infty$, respectively when $\varkappa\to\infty$, both these vacuum laws reduce to the
 \textbf{Maxwell}-\textbf{Lorentz} law $\bH = \bB \quad \&\ \quad \bE = \bD$.

 Given a subluminal velocity $|\dot\qV(t)| < c$, the field \textbf{Cauchy} problems are globally well-posed in the sense
of distributions for both the \textbf{ML} and \textbf{MBLTP} field equations.
  The \textbf{MBI} {field} Cauchy problem, unfortunately, has not yet been conquered to the extent needed. 
  {{Global well-posedness}} of the classical initial value problem has only been shown with {small data} (no charges!) in
\cite{Speck}; \textbf{F. Pasqualotto} presented an extension of \textbf{Speck}'s result to \textbf{MBI} field evolutions
on the \textbf{Schwarzschild} background \cite{FP}.
  A local well-posedness result for  \textbf{MBI} field evolutions with subluminal point sources (and inevitably large data)
should be possible, but so far only the special case of electrostatic solutions with $N$ point charge sources has been conquered
in \cite{mikiCMP}.
 There it was shown that a unique finite-energy electrostatic weak solution of the \textbf{MBI} field equations with $N$
point charges placed anywhere in $\Rset^3$ exists, and that the solution is real analytic away from the point charges for any choices of 
their signs and magnitudes.

 For the {field momentum densities} $\PiV$ (dropping the superscript ``field'') one has the following expressions.
 For the \textbf{ML} and for \textbf{MBI} field equations,
\begin{equation}
4\pi c\PiV = \bD\times \bB,
\end{equation}
whereas for the \textbf{MBLTP} field equations,
\begin{equation}
\textstyle
4\pi c\PiV
=
\bD\times\bB + \bE\times\bH - \bE\times\bB - \varkappa^{-2} \big(\nabla\cdot\bE\big)\big(\nabla\times\bB -\varkappa\, \dot\bE\big).
\end{equation}
 For typical 
\textbf{MBLTP} field evolutions with point sources we showed that $\PiV(t,\sV)$ is in 
$L^1_{loc}(\Rset^3)$, in particular about each $\qV(t)$, see \cite{mikishadi}.
 We expect such a result also for \textbf{MBI} fields.
 It is surely false in general for \textbf{ML} fields!
 
 With appropriate decay rates at spatial infinity imposed on the field initial data, 
the \textbf{MBLTP} field momentum $\int_{\Rset^3}\PiV(t,\sV)\drm^3s$ exists for all $t$. 
 Moreover, given \textbf{Lipschitz} maps $t\mapsto \qV(t)$, $t\mapsto\vV(t)$ and bounded $t\mapsto\aV(t)$, we showed that
$\Ddt\int_{\Rset^3}\PiV(t,\sV)\drm^3s$ exists for all $t$. 
 
 The crucial step in showing that \refeq{eq:momISconserved}, with \refeq{eq:MOMinTERMSofVELO}, yields an equation of motion is now
the following.
 The fields $\bB,\bD,\bE,\dot\bE$ (and $\bH$) at $(t,\sV)$ are given by explicit functionals of the vector functions
$\qV(\cdot)$ and $\vV(\cdot)$, and $\bD\ \&\ \bH$ also involve ${\ba}(\cdot)$;
their dependence on ${\ba}(\cdot)$ is linear.
 For $t<0$ we set $\qV(t)= \qV(0)+\vV(0)t$ and $\vV(t)=\vV(0)$, and $\aV(t)=\NullV$.
 Treating $\qV(\cdot)$ and $\vV(\cdot)$ as given, and ${\ba}(\cdot)$ as independent vector function variable for $t>0$, 
\refeq{eq:momISconserved} together with \refeq{eq:MOMinTERMSofVELO} 
is equivalent to a \textbf{Volterra} integral equation for $\ba = \ba[\qV,\pV]$, viz.
\begin{equation}\label{aVOLTERRAeqn}
{\ba}
= 
 W[\pV]\cdot \Big( 
\fV^{\mbox{\tiny{vac}}}[\qV,\vV]
+
\fV^{\mbox{\tiny{source}}}[\qV,\vV;{\ba}]\Big)
\end{equation}
where 
\begin{equation}
\vV = \frac{1}{\mbare} \frac{\pV}{\sqrt{1 +\frac{|\pV|^2}{\mbare^2 c^2 }}};\quad \mbare\neq 0
\end{equation}
and
\begin{equation}
W[\pV]
:= 
\sign(\mbare^{}) \frac{c}{\sqrt{\mbare^{2}c^2 + |\pV|^2}}
\left[\ID - \frac{\pV\otimes \pV}{{{\mbare^{2}c^2} + {|\pV|^2}}}\right],
\end{equation}
and where we have written the field as a sum of a source-free (vacuum) field and a field having the point charge as source, 
resulting in a \textbf{Lorentz} force due to that vacuum field,
\begin{equation}
\fV^{\mbox{\tiny{vac}}}[\qV,\vV](t) \equiv 
- e \left[ \bE^{\mbox{\tiny{vac}}}(t,\qV(t))+ \tfrac1c\vV(t)\times \bB^{\mbox{\tiny{vac}}}(t,\qV(t))  \right],
\end{equation}
and a ``self''-type force $\fV^{\mbox{\tiny{source}}}[\qV,\vV;{\ba}]$,  in \emph{\textbf{BLTP} electrodynamics}\footnote{None of 
the four original contributors formulated a well-defined expression for the force, yet we believe that our formulation
accomplishes what they had intended; hence the name of the theory.}
given by
\begin{alignat}{2}
\fV^{\mbox{\tiny{source}}}[\qV,\vV;{\ba}](t)
&  = - \frac{\drm}{\drm{t}} \displaystyle\int_{\Rset^3}\PiV^{\mbox{\tiny{source}}}(t,\sV)  d^3{s} 
\label{sourceforce}
\\
&  = - \frac{\drm}{\drm{t}} \displaystyle\int_{B_{ct}(\qV_0)}\left(\PiV^{\mbox{\tiny{source}}}(t,\sV) -\PiV^{\mbox{\tiny{source}}}(0,\sV-\qV_0-\vV_{\!0}t)\right) d^3{s} \\
&=  \tfrac{e^2} {4\pi } \biggl[ \biggr.
 - {\mathbf{Z}}_{\boldsymbol{\xi}}^{[2]}(t,t) + {\mathbf{Z}}_{\boldsymbol{\xi}^\circ}^{[2]}(t,t) \, 
\\ \notag
&   -\!\!\! \;{\textstyle\sum\limits_{0\leq k\leq 1}}\! c^{2-k}(2-k)\!\!
\displaystyle  \int_0^{t}\! 
\Bigl[{\mathbf{Z}}_{\boldsymbol{\xi}}^{[k]}\big(t,t^{\mathrm{r}}\big)
-\!
{\mathbf{Z}}_{\boldsymbol{\xi}^\circ}^{[k]}\big(t, t^{\mathrm{r}}\big)\Bigr]
(t- t^{\mbox{\tiny{r}}})^{1-k} \drm{t^{\mbox{\tiny{r}}}} 
\\ \notag
& -\!\!\!  \;{\textstyle\sum\limits_{0\leq k\leq 2}}\! c^{2-k}\!
\displaystyle  \int_0^{t}\! 
\Bigl[\tpddt{\mathbf{Z}}_{\boldsymbol{\xi}}^{[k]}\big(t,t^{\mathrm{r}}\big)
-\!
\tpddt{\mathbf{Z}}_{\boldsymbol{\xi}^\circ}^{[k]}\big(t, t^{\mathrm{r}}\big)\Bigr]
(t- t^{\mathrm{r}})^{2-k} \drm{t^{\mathrm{r}}}  \biggl. \biggr]
\end{alignat}
where $\boldsymbol{\xi}(t) \equiv (\qV,\vV,{\ba})(t)$, and 
$\boldsymbol{\xi}^\circ(t) \equiv (\qV_0+\vV_0t,\vV_0,{\boldsymbol{0}})$, and where
\begin{alignat}{1}
{\mathbf{Z}}_{\boldsymbol{\xi}}^{[k]}\big(t,t^{\mathrm{r}}\big) = 
 \displaystyle  \int_0^{2\pi}\!\! \int_0^{\pi}\!
\left(1-\tfrac1c\abs{\vV^{}(t^\mathrm{r})}\cos\vartheta\right)
 \boldsymbol{\pi}_{\boldsymbol{\xi}}^{[k]}\big(t,\qV(t^\mathrm{r}) + c(t-t^\mathrm{r})\nV 
\big) 
\sin\vartheta \drm{\vartheta}\drm{\varphi}\,,
 \end{alignat}
with $\nV =\left( \sin\vartheta \cos\varphi\ ,\ \sin\vartheta \sin\varphi\ ,\ \cos\vartheta \right)$, and where,
with the kernels
\begin{alignat}{1}
\mathrm{K}_{\boldsymbol{\xi}}(t',t,\sV) & :=
\tfrac{J_1\!\bigl(\varkappa\sqrt{c^2(t-t')^2-|\sV-\qV(t')|^2 }\bigr)}{\sqrt{c^2(t-t')^2-|\sV-\qV(t')|^2}^{\phantom{n}}},\\
\mathbf{K}_{\boldsymbol{\xi}}(t',t,\sV) & := 
\tfrac{J_2\!\bigl(\varkappa\sqrt{c^2(t-t')^2-|\sV-\qV(t')|^2 }\bigr)}{{c^2(t-t')^2-|\sV-\qV(t')|^2}^{\phantom{n}} }
 \left(\sV-\qV(t')- \vV(t')(t-t')\right), 
\end{alignat}
we have 
\begin{alignat}{1}
\hspace{-0.5truecm}
 \boldsymbol{\pi}_{\boldsymbol{\xi}}^{[0]}(t,\sV) =
& -
 \varkappa^4 \frac14\left[
{\textstyle{
\frac{ \left({\nV(\qV,\sV)}_{\phantom{!\!}}-\frac1c{\vV}\right)\crprd 
\left({\color{black} \frac1c{\vV}\crprd{\nV(\qV,\sV)} } \right) }{ 
\bigl({1-\frac1c {\vV}\cdot\nV(\qV,\sV)}\bigr)^{\!2} } }}\right]_{\mathrm{ret}}\\ \notag
&+ \varkappa^4\frac12\left[
{\textstyle{
\frac{ {\nV(\qV,\sV)}_{\phantom{!\!}}-\frac1c{\vV}}{ {1-\frac1c {\vV}\cdot\nV(\qV,\sV)} }
             }}\right]_{\mathrm{ret}} 
\crprd \!
\int_{-\infty}^{t^\mathrm{ret}_{\boldsymbol{\xi}}(t,\sV)}\!\!\!\!
{\vV(t')}\crprd \mathbf{K}_{\boldsymbol{\xi}}(t',t,\sV)\drm{t'} 
\\ \notag
& - \varkappa^4\frac12\left[{\textstyle{ 
\frac{ {\color{black} \frac1c{\vV}\crprd  {\nV(\qV,\sV)} } }{
      1-\frac1c {\vV}\cdot\nV(\qV,\sV)}}}\right]_{\mathrm{ret}} 
\crprd \int_{-\infty}^{t^\mathrm{ret}_{\boldsymbol{\xi}}(t,\sV)}\!\!\!\!
 c\mathbf{K}_{\boldsymbol{\xi}}(t',t,\sV)\drm{t'} 
\\ \notag
& - \varkappa^4 \int_{-\infty}^{t^\mathrm{ret}_{\boldsymbol{\xi}}(t,\sV)} \!\!\!\!
 c\mathbf{K}_{\boldsymbol{\xi}}(t',t,\sV)\drm{t'} \crprd \int_{-\infty}^{t^\mathrm{ret}_{\boldsymbol{\xi}}(t,\sV)} \!\!\!\!
{\vV(t')}\crprd \mathbf{K}_{\boldsymbol{\xi}}(t',t,\sV)\drm{t'} 
\\ \notag
& - \varkappa^4 c\int_{-\infty}^{t^\mathrm{ret}_{\boldsymbol{\xi}}(t,\sV)} \!\!\!\! \mathrm{K}_{\boldsymbol{\xi}}(t',t,\sV)\drm{t'} 
 \int_{-\infty}^{t^\mathrm{ret}_{\boldsymbol{\xi}}(t,\sV)}  \!\!\!\!\mathrm{K}_{\boldsymbol{\xi}}(t',t,\sV) {\vV}(t')\drm{t'} ,
\end{alignat}
\begin{alignat}{1}
\hspace{-0.5truecm}
 \boldsymbol{\pi}_{\boldsymbol{\xi}}^{[1]}(t,\sV) = & 
- \varkappa^2 
\left[
{\textstyle{
{\color{black}
{\nV(\qV,\sV)}\frac{\left({\nV(\qV,\sV)}\crprd [{ \left({\nV(\qV,\sV)}-\frac1c{\vV}\right)\crprd \aV }]\right)\cdot
\frac1c\vV}{ c^2 \bigl({1-\frac1c {\vV}\cdot\nV(\qV,\sV)}\bigr)^{\!4} }    }
           }}
+
{\textstyle{
{\nV(\qV,\sV)}_{\phantom{!\!}}\crprd\frac{ \left({\nV(\qV,\sV)}_{\phantom{!\!}}-\frac1c{\vV}\right)\crprd {\aV} }{
      2 c^2 \bigl({1-\frac1c {\vV}\cdot\nV(\qV,\sV)}\bigr)^{\!3} }
}}\right]_{\mathrm{ret}}\\ \notag
&- \varkappa^2\left[
{\textstyle{
{\nV(\qV,\sV)}_{\phantom{!\!}}\crprd\frac{ \left({\nV(\qV,\sV)}_{\phantom{!\!}}-\frac1c{\vV}\right)\crprd {\aV} }{
      c^2 \bigl({1-\frac1c {\vV}\cdot\nV(\qV,\sV)}\bigr)^{\!3} }
}}\right]_{\mathrm{ret}} \!\!
\crprd \!
\int_{-\infty}^{t^\mathrm{ret}_{\boldsymbol{\xi}}(t,\sV)}\!\!\!\!
{\vV(t')}\crprd \mathbf{K}_{\boldsymbol{\xi}}(t',t,\sV)\drm{t'} 
\\ \notag
& + \varkappa^2\left[\nV(\qV,\sV)\crprd \biggl[{\textstyle{\nV(\qV,\sV)\crprd 
\frac{\left({\nV(\qV,\sV)}_{\phantom{!\!}}-\frac1c{\vV}\right)\crprd{\aV} }{
      c^2\bigl({1-\frac1c {\vV}\cdot\nV(\qV,\sV)}\bigr)^{\!3} }
}}\biggr]\right]_{\mathrm{ret}} \!\!\! 
\crprd\! \int_{-\infty}^{t^\mathrm{ret}_{\boldsymbol{\xi}}(t,\sV)} \!\!\!\!
 c\mathbf{K}_{\boldsymbol{\xi}}(t',t,\sV)\drm{t'} 
\\ \notag
& -\varkappa^3 
 \left[\textstyle\frac{1}{{1-\frac1c {\vV}\cdot\nV(\qV,\sV)} }\right]_{\mathrm{ret}} 
 \int_{-\infty}^{t^\mathrm{ret}_{\boldsymbol{\xi}}(t,\sV)}\!\!\!\!
 \mathrm{K}_{\boldsymbol{\xi}}(t',t,\sV)\left[{\vV}({t^\mathrm{ret}_{\boldsymbol{\xi}}(t,\sV)}))+{\vV}(t')\right]\drm{t'},
 \end{alignat}
\begin{alignat}{1}
\hspace{-0.5truecm}
\boldsymbol{\pi}_{\boldsymbol{\xi}}^{[2]}(t,s) = & - \varkappa^2
\left[\textstyle\frac{1}{\bigl({1-\frac1c {\vV}\cdot\nV(\qV,\sV)}\bigr)^{\!2} }\frac1c{\vV}
{\color{black}- \Big[\!{1-\tfrac{1}{c^2}\big|\vV_n\big|^2}\!\Big]
\frac{ \left({\nV(\qV,\sV)}_{\phantom{!\!}}-\frac1c{\vV}\right) \crprd \left(\frac1c\vV\crprd \nV(\qV,\sV)\right) }{
      \bigl({1-\frac1c {\vV}\cdot\nV(\qV,\sV)}\bigr)^4 }}
\right]_{\mathrm{ret}} 
\\ \notag
&  +\varkappa^2 \left[\Big[\!{1-\tfrac{1}{c^2}\big|\vV\big|^2}\!\Big]\nV(\qV,\sV)\crprd {\textstyle{
\frac{ {\nV(\qV,\sV)}_{\phantom{!\!}}-\frac1c{\vV} }{
      \bigl({1-\frac1c {\vV}\cdot\nV(\qV,\sV)}\bigr)^{\!3} }
}}\right]_{\mathrm{ret}} \crprd\int_{-\infty}^{t^\mathrm{ret}_{\boldsymbol{\xi}}(t,\sV)} \!\!\!\!
 c\mathbf{K}_{\boldsymbol{\xi}}(t',t,\sV)\drm{t'} \\ \notag
& -  \varkappa^2\left[\Big[\!{1-\tfrac{1}{c^2}\big|\vV\big|^2}\!\Big]
{\textstyle{
\frac{ {\nV(\qV,\sV)}_{\phantom{!\!}}-\frac1c{\vV} }{
      \bigl({1-\frac1c {\vV}\cdot\nV(\qV,\sV)}\bigr)^{\!3} }
}}\right]_{\mathrm{ret}} 
\crprd \!
\int_{-\infty}^{t^\mathrm{ret}_{\boldsymbol{\xi}}(t,\sV)} \!\!\!\!
{\vV(t')}\crprd \mathbf{K}_{\boldsymbol{\xi}}(t',t,\sV)\drm{t'} ,
\end{alignat}
and $\big|_\mathrm{ret}$ means that $\qV(\tilde{t})$, $\vV(\tilde{t})$, ${\aV}(\tilde{t})$ 
are evaluated at~$\tilde{t} = {t^\mathrm{ret}_{\boldsymbol{\xi}}}(t,\sV)$.\hspace{-1truecm}

\begin{rem}
 The decomposition of the electromagnetic fields into a sum of two types of fields, one with the point charge as source, the other
source-free, is to some extent arbitrary. For this reason 
\ul{it is futile to try to identify \emph{the self-field force} of a point charge.}
 At best one can speak of a ``self''-field force, the scare quotes referring to the ambiguity in identifying how much of the field
 is ``self''-generated by the charge and how much is not.
\end{rem}

The following \emph{key proposition} about the \textbf{Volterra} equation is proved in  \cite{mikishadi}.
Its proof takes several dozen pages of careful estimates.

\noindent
\begin{prp}\label{prop}
 Given $C^{0,1}$ maps ${t\mapsto{\qB}(t)}$ and $t\mapsto{\pB}(t)$, with Lip$(\qB)=v$, Lip$(\vB)=a$ big enough, and
$|\vB(t)|\leq v<c$, 
the \textbf{Volterra} equation \refeq{aVOLTERRAeqn} as a fixed point map has a unique $C^0$ solution 
$t\mapsto\ba(t) = \boldsymbol{\alpha}[\qB(\,\cdot\,),\pB(\,\cdot\,)](t)$ for $t\geq 0$. 
 Moreover, the solution depends \textbf{Lipschitz}-continuously on the maps
$t\mapsto{\qB}(t)$ and $t\mapsto{\pB}(t)$ (treated as independent). 
\end{prp}

  The well-posedness result of the joint initial value problem for \textbf{MBLTP} fields and their point charge
sources is a corollary of Proposition \ref{prop}.
 Namely, now setting $\aV(t):=\Ddt\vV(t)$ and $\vV(t):=\Ddt\qV(t)$, and recalling \refeq{eq:MOMinTERMSofVELO}, the solution to
the \textbf{Volterra} integral equation for the acceleration poses a \textbf{Newton}-type second-order initial value problem 
for the position of the point charge with a complicated yet \textbf{Lipschitz}-continuous force. 
 Once the motion is computed, inserting the vector functions $\qV(\cdot)$, $\vV(\cdot)$, and ${\ba}(\cdot)$ of the solution
into the functionals of the fields yields $\bB,\bD,\bE,\dot\bE$ at $(t,\sV)$ for $t>0$, too.
 This is a Theorem in \cite{mikishadi}, summarized informally as follows.

\noindent
\begin{thm}
 As a consequence of Proposition \ref{prop}, the joint initial value problem for \textbf{MBLTP} fields and their point charge source
is equivalent to the fixed point equations 
\begin{alignat}{2}
&\qV(t) = \qV(0) + \frac{{\textstyle{1}}}{\mbare} 
\int_0^t  \frac{\pV}{\sqrt{1 + {|\pV|^2}/{\mbare^{2} }}}(\tilde{t})d{\tilde{t}}  
& =: Q_t(\qV(\cdot),\pV(\cdot)) \\
& \pV(t) = \pV(0) -  \displaystyle\int_{\Rset^3}\left(\PiV(t,\sV) -\PiV(0,\sV)\right) d^3{s} 
& =:P_t(\qV(\cdot),\pV(\cdot)),
\end{alignat}
where $Q_t$ and $P_t$ are \textbf{Lipschitz} maps.
 Thus, \textbf{BLTP} electrodynamics is locally well-posed.
\end{thm}

 In fact, in \cite{mikishadi} the \textbf{Cauchy} problem for the \textbf{MBLTP} field with $N$ point charges is treated.
 \emph{Local well-posedness} is proved for \emph{admissible initial data} (see below) \&\ ${\mbare}\neq 0$, and 
\emph{global well-posedness} shown to hold if {in finite time}: 

(a) no particle reaches the speed of light;

(b) no particle is infinitely accelerated;

(c) no two particles reach the same location.

\noindent
By ``admissible'' initial data we mean the following: the initial particle velocities are subluminal ($|\vV(0)|<c$) and no
two particles occupy the same location; the electromagnetic initial fields are the sum of a spatially sufficiently
rapidly decaying vacuum field plus $N$ fields each with a single point charge source --- the sourced fields are 
boosted electrostatic fields with boost velocity equal to the initial velocity of the source. 

 Although our result seems to be the first formulation of a well-posed joint initial value problem for classical electromagnetic 
fields and their point charge sources, and this endows \textbf{BLTP} electrodynamics with a mathematically superior status compared
to the ill-defined \textbf{Lorentz} electrodynamics, we do not claim that \textbf{BLTP} electrodynamics is the correct 
classical limit of the elusive mathematically well-formulated quantum theory of electromagnetism.
 In particular, the \textbf{MBLTP} field equations feature ``physical'' {oddities}: 

(o1) {a field energy functional which is unbounded below}; 

(o2) {subluminal transversal electromagnetic wave modes};

(o3) {longitudinal electrical wave modes}.

\noindent
Moreover, since the \textbf{MBLTP} field equations demand initial data for $\bB,\bD,\bE,\dot\bE$ at $t=0$ while physically we only
can prescribe $\bB$, $\bD$ (N.B.: in \textbf{Lorentz} electrodynamics, $\bD=\bE$; not so in \textbf{BLTP} electrodynamics), one
needs to find a prescription which expresses $\bE,\dot\bE$ at $t=0$ in terms of $\bB,\bD$ at $t=0$.
 In \cite{mikishadi} we show that a reasonable choice is the map
$(\bB,\bD)_0 \mapsto (\bE,\dot\bE)_0$ obtained by \emph{maximizing the field energy} w.r.t. $\bE_0$ and $\dot\bE_0$ (treated 
independently).
 This can be made co-variant by stipulating that the maximization is carried out in the \textbf{Lorentz} frame 
in which the total particle momentum vanishes initially.

 Back to the mathematically superior status of \textbf{BLTP} electrodynamics, one can now apply rigorous analysis, and 
controlled numerical techniques, to study the theory. 
 For instance, a rigorous comparison of our expression \refeq{sourceforce} for the {``self''-force} with 
a differently defined ``self''-field force which was studied in \cite{GratusETal} 
was carried out recently by Hoang \&\ Radosz and their students, see \cite{VuETC}, \cite{HoangRadosz}.
 One of our next projects is to rigorously extract effective equations of motion with more user-friendly 
expressions for the ``self''-force. 
 In particular, whether or to which extent the \textbf{Landau--Lifshitz} equation approximately governs the motion is 
an interesting question.

                \subsubsection{Topologically non-trivial flat spacetimes with time-like singularities}

 The {\em zero-gravity limit} of singular spacetimes does not automatically yield \textbf{Minkowski} spacetime minus a number of
world-lines. 
 In \cite{TZ2014} the zero-$G$ limit of \textbf{Carter}'s maximal analytical extension of the electromagnetic \textbf{Kerr--Newmann} 
spacetime was analyzed. 
  The limiting spacetime  is axially symmetric and static.
  It is locally isometric to \textbf{Minkowski} spacetime, but is topologically non-trivial, featuring \textbf{Zipoy} topology:
  its constant time slices are double-sheeted, and have the topology of $\mathbb{R}^3$ branched over the un-knot.
 The spacetime is singular on a 2-dimensional time-like cylinder $\Sset^1\times \Rset$, which is the world-tube of a 
space-like ring-type singularity. 
 The most intriguing aspect of this solution is that at any instant of time the ring singularity, when viewed from one of the two sheets of 
space, appears to be positively charged, and from the other sheet, negatively charged, as first noticed by \textbf{Carter}.
 The electromagnetic fields it supports were discovered in the 19th century by \textbf{P. Appell} as ``multi-valued electromagnetic
fields,'' while \textbf{A. Sommerfeld} soon realized that they represent single-valued electromagnetic fields on a topologically 
non-trivial space.

 The next natural step in this direction would be to formulate the corresponding zero-gravity two-body problem with 
two space-like ring-type singularities analogous to z$G$\textbf{KN} evolving in time jointly with the electromagnetic fields they generate.
 We know for example that the space should be four-sheeted (in general, $2^N$ sheets are needed for $N$ ring-type singularities).
 However, as the z$G$\textbf{KN} fields solve the \textbf{Maxwell--Lorentz} vacuum field equations away from the singularities, it 
is not surprising that one again encounters the infinite field energy-momentum problems which plague \textbf{Lorentz} electrodynamics. 
 Thus one would first need to find either \textbf{MBI} or \textbf{MBLTP} analogues of the \textbf{Appell--Sommerfeld} fields.
 The nonlinearity of the \textbf{MBI} field equations makes this a daunting task, but we are confident that the feat can be
accomplished with the \textbf{MBLTP} field equations.
 In that case the zero-$G$ \textbf{EIH}-type approach explained in the previous subsection should also allow the
formulation of a well-posed initial value problem for ring-type singularities and their electromagnetic \textbf{MBLTP} fields.
 Note that the law of motion would most likely be a system of partial differential equations, as the ring-type singularity has to
be allowed to bend, twist, warp, and stretch.

                \subsection{Turning on gravity: $G>0$}

                \subsubsection{The neighborhood of Minkowski spacetime}

 By rigorously establishing well-posedness of \textbf{BLTP} electrodynamics with point charges as a consequence of postulating 
the conservation law \refeq{eq:MOMconservation} for the total momentum vector-density \refeq{eq:totalMOMENTUMdef}, 
with the expressions for the particles given by the usual special-relativistic ones and those for the fields determined
by the field equations, \cite{mikishadi} demonstrates that a key idea of the 1938 \textbf{EIH} 
paper is viable in the zero-$G$ limit when applied with a suitable set of electromagnetic field equations,
and with non-zero bare rest mass assigned to the point charges.
 We now address the question whether this result extends continuously to a $G>0$ neighborhood of special relativity,
free of black holes. 
 By a remark of \textbf{Geroch--Traschen} \cite{GerochTraschen} a {no}-Black-Holes spacetime with a one-dimensional time-like singularity 
cannot exist if the singularity has {positive} {bare mass}, different
from the special-relativistic case, where the bare mass merely had to be non-zero.

 One of the main ingredients of the \textbf{EIH} argument is of course, that the conservation law of energy-momentum is implied by the
twice contracted 2$^{nd}$ \textbf{Bianchi} identity in concert with \textbf{Einstein}'s general-relativistic field equations.
 However, for the spacetimes with time-like singularities envisaged by \textbf{EIH} (and \textbf{Weyl})
this is not automatically true.
 An important step, therefore, is to determine the \textbf{Lorentz} spacetimes with time-like singularities on which 
the twice contracted 2$^{nd}$ \textbf{Bianchi} identity holds in a {weak form}. 

 Together with \textbf{A. Y. Burtscher} and \textbf{J. Stalker} we have begun a systematic study of the favorable conditions.
 In \cite{BKSTZ} we study the simplest non-trivial case: static spherically symmetric spacetimes with a single time-like singularity.
 We obtain some necessary and, for certain electromagnetic vacuum laws, also sufficient conditions that
the twice contracted 2$^{nd}$ \textbf{Bianchi} identity holds in a {weak form}. 
 Interestingly, in the naked singularity regime (no Black Hole!), the answer is \emph{negative} for the electromagnetic 
\textbf{Reissner--Weyl--Nordstr\"om} spacetime, but \emph{positive} for the \textbf{Hoffmann} spacetime\footnote{It is curious
 that although \textbf{Hoffmann} had worked out his spacetime solution by 1935 \cite{Hoffmann}, 
 the fact that his spacetime with non-positive bare mass is less singular than the \textbf{RWN} spacetime
 did not in 1938 seem to suggest to \textbf{EIH} to use the \textbf{MBI} instead of \textbf{Maxwell--Lorentz} field theory.}
in which electromagnetic \textbf{MBI} fields are coupled with \textbf{Einstein}'s gravity; see \cite{TZ} for a rigorous discussion.
 Our goal is to generalize our study,  one step at a time, to dynamical spacetimes without symmetry.

 The crucial question, then, is whether the weak twice contracted 2$^{nd}$ \textbf{Bianchi} identity implies the law of
the electromagnetic spacetime's time-like singularities with bare energy-momentum assigned to them.
 Interestingly, it seems that merely changing \textbf{Maxwell}'s into the \textbf{BLTP} or \textbf{MBI} vacuum law is by 
itself not sufficient to allow the formulation of a well-defined joint initial value problem for the massive point charges 
and the electromagnetic and gravitational fields they generate:
the gravitational coupling also needs to be changed, for instance in the manner of \textbf{Jordan, Brans--Dicke}, or 
$f(\tenseur{R})$ gravity, to obtain a well-defined joint \textbf{Cauchy} problem. 
 Put differently, the gravitational coupling of spacetime structure with bare matter and the electromagnetic fields needs to be
``mediated'' by a certain type of scalar field which moderates the strength of the spacetime singularities enough so that 
the strategy explained in the previous subsection can be applied. 
 This scalar field in the classical theory would play a role vaguely reminiscent of the role played by the 
scalar Higgs field in the quantum field-theoretical standard model of elementary particle physics.
 
                \subsubsection{The neighborhood of z$G$KN-type spacetimes}

 Everything stated in the previous subsection about the neighborhood of the \textbf{Minkowski} spacetime has an analogue
problem for the gravitational neighborhood of topologically non-trivial flat spacetimes of (generalized) z$G$\textbf{KN} type. 
 
\bigskip
\noindent
\textbf{Acknowledgement}:
We gratefully acknowledge interesting discussions with: E. Amorim, A. Burtscher, H. Carley, D. Deckert, V. Hartenstein,
V. Hoang, M. Kunze, V. Perlick, M. Radosz, J. Speck, H. Spohn. We thank the organizers, R. Ruffini and R. Jantzen, 
for inviting us to MG15.

\medskip
\noindent
\textbf{Note added:} It was pointed out to me that what I call the Landau--Lifshitz equation of motion should rather
be called the Eliezer--Ford--O'Connell equation of motion.

%

\end{document}